\documentstyle{elsartwb}

\input epsf.sty

\begin{document}

\begin{frontmatter}
  \title{\bf Description of the particle ratios and transverse-momentum spectra
    for various centralities at RHIC in a single-freeze-out
    model\thanksref{grant}} \thanks[grant]{Research supported in part
    by the Polish State Committee for Scientific Research, grants
    2~P03B~09419 and 2~P03B~11623}
\thanks[emails]{%
\hspace{0mm} E-mail: b4bronio@cyf-kr.edu.pl, florkows@amun.ifj.edu.pl}
\author[INP]{Anna Baran}, 
\author[INP]{Wojciech Broniowski}, and
\author[INP,Kielce]{Wojciech Florkowski}
\address[INP]{The H. Niewodnicza\'nski Institute of Nuclear Physics,
        ul. Radzikowskiego 152,  PL-31342 Krak\'ow, Poland}
\address[Kielce]{Institute of Physics, \'Swi\c{e}tokrzyska Academy,
ul.~\'Swi\c{e}tokrzyska 15, PL-25406~Kielce,~Poland}

\begin{abstract}
  The single-freeze-out model of hadron production is used to describe
  the particle ratios and the transverse-momentum spectra from RHIC.
  The emphasis is put on the new measurements done at the highest beam
  energy of $\sqrt{s_{NN}}$=200 GeV. An overall very good agreement is
  found between the data and the model predictions. The data for
  different centrality windows are analyzed separately. A simple
  scaling of the two expansion parameters of the model with the
  centrality is found. Interestingly, this scaling turns out to be
  equivalent to the scaling of hadron production with the number of
  wounded nucleons.
\end{abstract}


\end{frontmatter}
\vspace{-7mm} PACS: 25.75.Dw, 21.65.+f, 14.40.-n

\section{Introduction}

Ultra-relativistic heavy-ion collisions offer a unique possibility to
produce hot and dense hadronic matter in laboratory conditions.  With
the new results presently coming from RHIC \cite{qm01,qm02,hir02} we
have entered an exciting era when various model predictions and
theoretical ideas may be confronted directly with the large amount of
diverse and accurate data. One of the key issues in the field is the
question of the degree of thermalization of the hadronic matter
produced in heavy-ion collisions. From the outset of the heavy-ion
physics the thermal models have been useful and successful to a large
degree in the description of data
\cite{RLT,BM12,Cleymans,BM3,gazgor,YenGor,finland,BM4,wfwbmm,bialasqm,%
rafqm,kochqm,prorok,huovin}
(for a recent review see \cite{review}).

In this paper we analyze the hadron production at mid-rapidity
measured at RHIC at the full beam energy of $\sqrt{s_{NN}}$=200~GeV
within the {\em single-freeze-out model} of Ref.
\cite{wbwf,str,zakop}.  The new element of this work is the
application of the approach to data at various values of the {\em
  centrality} parameter, $c$ (or the impact parameter, $b$), which
allows us for a closer look at the geometry and flow of the hadronic
system formed in non-central collisions. The obtained systematics for
the geometric parameters as functions of $c$ is very interesting, as
it conforms to the wounded-nucleon scaling
\cite{wounded,budzan,woundedphobos}. We find that the invariant time
at freezeout, $\tau$, and the transverse size, $\rho_{\rm max}$,
behave in a very systematic way for not-too-large values of $c$. In
particular $\rho_{\rm max} \sim (1-c)$, and $\tau/\rho_{\rm max}\simeq {\rm
  const}$.

The single-freeze-out model was formulated in Ref.
\cite{wbwf,str,zakop} where it was successfully used to describe the
transverse-momentum spectra of hadrons measured at the then-available
lower beam energy of $\sqrt{s_{NN}}$=130 GeV. The model combines a
typical thermal approach, used to study the ratios of hadron
abundances \cite{RLT,BM12,Cleymans,BM3,gazgor,YenGor,finland,BM4} with
a hydrodynamic expansion including in a natural way the longitudinal
and transverse flow.  A characteristic and important feature of the
model is the complete treatment of the resonances in both the
calculation of the ratios of hadron yields and the analysis of the
spectra. The model and the data at $\sqrt{s_{NN}}$=130 GeV, including strange-quark
particles such as $\phi$, $\Lambda$, $\Xi$, $\Omega$, as well as
$K^\ast$, have been found to be in a surprising agreement \cite{str}.
A comparison of the model predictions with the data for particle
abundances and transverse-momentum spectra, collected at the maximum
RHIC energy of 200~GeV and presented in detail in this paper, bring
further evidence for rapid thermalization, or a statistical nature, of
the hadronic matter produced at mid-rapidity in heavy-ion collisions.
Moreover, with our simple approach we form a basis for the explanation
of most of the soft features of the hadron production observed at
RHIC.

The paper is organized as follows: In the next Section we outline the
main assumptions of the model, bringing up for completeness some of
the basic material of Refs.~\cite{wbwf,str,zakop}. In Sect. 3 we study
the ratios of hadron yields at $\sqrt{s_{NN}}=200$~GeV and find the
thermodynamic parameters characterizing the freeze-out. In Sect. 4 we
calculate the transverse-momentum spectra.  The calculations are done
for different centrality windows, such as defined by the experiment.
Naturally, since the geometry of the hot system formed in the
collision depends on the centrality, we find a dependence of the
geometric/expansion parameters on $c$. Interestingly, this dependence
reflects quite accurately, for not too large $c$, the scaling of
hadron production with the number of wounded nucleons.

\section{The single-freeze-out model}

In our previous work \cite{wfwbmm,wbwf,str,zakop} we have shown that with the
complete treatment of hadronic resonances, the distinction between the 
traditionally considered two
freeze-outs, the chemical and the thermal (or kinetic) one (see the following discussion),
is not necessary. At least for the RHIC energies one can achieve a
very good explanation of the soft part of the hadronic data by assuming
a single freeze-out which takes place at the universal temperature
\begin{equation}
T_{\rm chem}=T_{\rm kin} \equiv T. \label{Tuniv}
\end{equation}

Clearly, a theoretical description of the whole duration of the
collision process of the two ultra-rela\-tivis\-tic heavy ions is
difficult, since different degrees of freedom are important at various
stages of the evolution.  Consequently, one is forced to describe each
stage within a different theoretical framework. In our approach we use
only hadronic degrees of freedom and thus concentrate on the latest
stages of the evolution of the system, namely when the hadrons, at
some instant, cease to interact and stream freely to the detectors.
This moment is called the freeze-out. Admittedly, this is a
far-reaching simplification. Generally speaking, the freeze-out in
itself may be a complicated process, involving duration in time, and a
hierarchy where different kinds of particles and different reactions
switch off at different times.  In general, by kinetic arguments, one
expects that reactions with lower cross sections switch off at higher
densities/temperatures, while those with larger cross sections last longer. Since,
in most cases, the elastic cross sections are larger than the
inelastic ones (a counter example, though, is the $p \bar p$ interaction),
one expects that the inelastic (or chemical) freeze-out occurs earlier
than the thermal freeze-out. On these grounds one may also argue that
strange (or charmed \cite{gazdz}) particles decouple earlier than
other hadrons.  With such a picture in mind one
may ``separate'' the freeze-out process into a series of more specific
freeze-outs associated to particular reaction channels. In fact, the
works of Heinz and collaborators \cite{heinzr} introduced the concept
of the chemical and the thermal (or kinetic) freeze-outs, with $T_{\rm
  chem} > T_{\rm kin}$.  At the chemical freeze-out the inelastic
interactions between the hadrons cease, and the chemical composition
of the system is fixed. Later, only the elastic interactions are
effective, leading to further cooling of the system.  At the thermal
freeze-out even the elastic interactions become ineffective and the
hadrons are completely decoupled from each other.

One should bare in mind, however, that the scales in the system
responsible for the thermalization and expansion are of similar order.
To be more precise, for expanding systems the criterion for the
thermal freeze-out to occur is \cite{hushu} 
\begin{equation}
\xi=\tau _{\exp }/\tau_{coll} \sim 1, \label{xi}
\end{equation} 
with
\begin{equation}
\tau _{\exp }=1/(\partial _{\mu }u^{\mu }).  \label{hubblecoll}
\end{equation}
With our parameterization for the expansion profile we obtain $\tau
_{\exp }\simeq 10\, \rm{GeV}^{-1}$, while Ref.~\cite{hushu} gives $\tau
_{coll} \simeq 10\, \rm{GeV}^{-1}$ at $T \simeq 165 \rm{\,MeV}$.  Thus the
discussion on the validity of our approach (early thermal freeze-out)
as opposed to approaches having later thermal freeze-out becomes a
discussion concerning the precise value of the factor in the criterion of Eq.~%
(\ref{xi}). It is possible that already at the moment of
hadronization the system is so dilute and possesses such a strong flow
that not only inelastic, but even the elastic processes can be
neglected. Thus our main assumption of Eq.~(\ref{Tuniv}) may be, in
the least, a good starting point.  At any case, assumptions of a
phenomenological model like ours are verified a posteriori by the
experiment. As we have shown in Refs.~\cite{wbwf,str,zakop} and
will also demonstrate in this work for the full RHIC energy of
$\sqrt{s_{NN}}=200$~GeV, we are in a rather comfortable situation.
The single-freeze-out hypothesis, when combined with expansion, offers
a very economic description of the data, with very few parameters (two
universal thermal parameters and two geometric parameters for each
centrality bin).

An experimental argument in favor of  a short time between the
chemical and kinetic freeze-outs has been provided by the observation
of resonances, such as the $K^\ast$ or $\rho$, by the STAR
Collaboration \cite{patricia,starKstar,starKstar2}, in the
invariant-mass spectra.  In the thermal approach the abundance of
these resonances is compatible with a high freeze-out temperature,
around 160~MeV. On the other hand, if the elastic scattering processes
($K-\pi$ for the $K^\ast$ or $\pi-\pi$ for the $\rho$) took place for
a much longer time, such that the system would cool off significantly,
the abundance of the resonances would be much lower.  We should point
out that the scenario of a single freeze-out, put forward in
Ref.~\cite{wbwf}, is compatible with the sudden hadronization scenario
of Ref.~\cite{suddenhad}.

Thus, our starting point is the hypothesis (\ref{Tuniv}).
Similarly to other thermal models of hadron production, we fix the
value of $T$, and also the value of the baryon chemical potential
$\mu_B$, by fitting the ratios of hadron abundances. The results of
such an analysis will be presented in the next Section.

The assumption of the single freezeout allows us to calculate
uniformly, {\em i.e.} within the same model, both the ratios of hadron
abundances and the trans\-verse-momentum spectra. In order to do this,
however, we need to make a choice of the freeze-out hypersurface ({\em
  i.e.} a three-dimensional volume in the four-dimensional space-time)
and of the four-velocity field of expansion at the freeze-out. A
priori, since we do not consider dynamics of the earlier stages here
and only make an educated guess, many choices are possible and for
each particular case one may check the validity of the proposed form
of the hypersurface and flow by confronting the output of the model
calculations to the data. Our choice has been made in the spirit of
the hydrodynamic calculations of
Ref.~\cite{bjorken,baym,Kolya,siemens,SSH,BL,cs1,Rischke,SH,cs2}, and is
defined by the condition
\begin{equation}
\tau = \sqrt{t^2-r^2_x-r^2_y-r^2_z} = {\rm const}.
\label{tau}
\end{equation}
To make the transverse size of the fire-cylinder,
\begin{equation}
\rho=\sqrt{r_x^2+r_y^2}, 
\label{rhodef}
\end{equation}
finite, we impose the condition 
\begin{equation}
\rho < \rho_{\rm max}. 
\label{rhomax}
\end{equation}
In addition, we assume that the four-velocity of the hydrodynamic
expansion at freeze-out is proportional to the coordinate (Hubble-like
expansion),
\begin{equation}
u^{\mu } =\frac{x^{\mu }}{\tau }=\frac{t}{\tau }\left(
1,\frac{r_{x}}{t},\frac{r_{y}}{t},\frac{r_{z}}{t}\right).
\label{umu}
\end{equation}
Such a form of the flow at freeze-out, as well as the fact that $t$
and $r_z$ coordinates are not limited and appear in the
boost-invariant combination in Eq. (\ref{tau}), imply that our model
is boost-invariant. In view of the recent data delivered by BRAHMS
\cite{brahmsy}, this assumption turns out to be justified for the
description of particle production in the rapidity range $-1<y<1$,
where the variation of the observed particle multiplicities is
moderate.

We note that the freeze-out hypersurface and the flow are jointly
controlled by the parameters $\tau$ and $\rho_{\rm max}$. Hence, our
model has altogether only four parameters (dependent on the colliding
energy and the centrality parameter), namely, the two thermodynamic
parameters, $T$ and $\mu_B$, and the two expansion (geometric)
parameters, $\tau$ and $\rho_{\rm max}$.  The thermodynamic parameters
turn out to depend on the centrality bin very weakly, reflecting the
very weak observed dependence of the particle ratios. On the other
hand, the geometric parameters do depend on $c$ and their dependence,
obtained from independent fitting, assumes a very intuitive form ({\em
  cf.} Sect.~6).

Other choices for the freezeout hypersurface and expansion have been
investigated in the literature \cite{lee}, with a recent study made in
Ref.~\cite{RafTorrieri}, where a class of parameterizations has been
tested with a limited number of resonances included.

\begin{table}[b]
\begin{centering}
\begin{tabular}{|l|r|c|}
\hline
& Model & Experiment \\ \hline \hline
\multicolumn{3}{|l|}{Fitted thermal parameters}\\ \hline\hline
$T$ [MeV] & 165.6$\pm 4.5$ &  \\ \hline
$\mu _{B}$ [MeV] & \ 28.5$\pm 3.7$ &  \\ \hline
$\mu _{S}$ [MeV] & \ \ \ \ \ 6.9 &  \\ \hline
$\mu _{I}$ [MeV] & \ \ \ \ \ -0.9 &  \\ \hline
$\chi ^{2}/n$ & 0.2 &  \\ \hline \hline
\multicolumn{3}{|l|}{Ratios used in the thermal analysis}\\ \hline\hline
$\pi ^{-}/\pi ^{+}$ & $1.009 \pm 0.003$ & 
\begin{tabular}{c}
$1.025\pm 0.006\pm 0.018$ \,\cite{phobos200} (0 - 12\%) \\   
$1.02 \pm 0.02 \pm 0.10 $ \,\cite{phenix200} (0 - 5\%)
\end{tabular}
\\ \hline 
$K^{-}/K^{+}$ & $0.939 \pm 0.008$ & 
\begin{tabular}{c}
$0.95\pm 0.03  \pm 0.03$ \,\cite{phobos200} (0 - 12\%) \\
$0.92 \pm 0.03 \pm 0.10$ \,\cite{phenix200} (0 - 5\%)
\end{tabular}
\\ \hline
${\overline{p}}/p$ & $0.74 \pm 0.04$ & 
\begin{tabular}{c}
$0.73\pm 0.02\pm 0.03$   \,\cite{phobos200} (0 - 12\%) \\
$0.70 \pm 0.04 \pm 0.10$ \,\cite{phenix200} (0 - 5\%) \\
$0.78 \pm 0.05$ \,\cite{vanBuren} (0 - 5\%)
\end{tabular}
\\ \hline 
${\overline{p}}/\pi^-$ & $0.104 \pm 0.010$ & $0.083 \pm 0.015$ \,\cite{star200}  
(0 - 5\%)\\ \hline
$K^-/\pi^-$ & $0.174 \pm 0.001$  & $0.156 \pm 0.020$ \cite{star200} (0 - 5\%) \\ \hline
$\Omega/h^-\,\,\, \times 10^3$ & $0.990 \pm 0.120$ & $0.887 \pm 0.111 \pm 0.133$ 
\,\cite{suire} (0 - 10\%)\\ \hline
${\overline \Omega}/h^- \,\,\, \times 10^3$ & $0.900 \pm 0.124$ & $0.935 \pm 0.105 
\pm 0.140$ 
\,\cite{suire} (0 - 10\%) \\ \hline
\end{tabular}
\caption{The model fit to the particle ratios measured at RHIC at $\sqrt{s_{NN}}=
  200$~GeV. Only stable hadrons (with respect to the strong
  interactions) are included.  The experimental and the theoretical
  pion yields are corrected for the weak decays
  \cite{patricia,star200}. The values in the brackets denote the centrality class.
The errors in the model predictions reflect the full errors of the 
experimental data points used in the $\chi^2$ method. Note the very good quality of the fit, reflected by 
the low value of $\chi^2$ per degree of freedom.}
\end{centering}
\label{tab:fitnores}
\end{table}

All resonances from the Particle Data Tables \cite{PDG} have been
included in our study.  The method and necessary formulas that
describe the decays in cascades can be found in Ref.~\cite{zakop}. We
stress the important role of including the high-lying states, which
although thermally suppressed, are increasingly numerous according to
the Hagedorn hypothesis \cite{hagedorn,myhag,bled,rafhag}. Moreover,
the inclusion of resonances is crucial for the description of the
$p_\perp$-spectra, as it increases the logarithmic slope as if
lowering the temperature \cite{wfwbmm}. This observation works in
favor of the hypothesis (\ref{Tuniv}).

\section{Particle ratios}

For boost-invariant systems the ratios of hadron abundances at
mid-rapidity are equal to the ratios of hadron densities (for a more
detailed discussion of this point, which includes the effects of the
resonance decays, see Ref. \cite{zakop}). Using this property, we can
determine thermodynamic parameters of our model directly from the
study of the ratios of hadron abundances.  We stress that we have only
two independent thermodynamic parameters in our model: the
temperature, $T$, and the baryon chemical potential, $\mu_B$. The
other two chemical potentials ($\mu_S, \,\mu_I$) are fixed by the
conditions that the initial strangeness of the system is zero and that the
ratio of the baryon number to the electric charge is the same as in
the colliding nuclei. This assumption could be lifted at the expense 
of carrying an extra parameter. However, in practice this is irrelevant, since 
the value of $\mu_I$ turns out to be
very small and the effects of the isospin violation can be neglected.

In Table 1 we show the results of our fit to the RHIC data collected
at $\sqrt{s_{NN}}$= 200 GeV. We include here only the stable hadrons
(with respect to strong interactions) and use the most central data.
The values of $T$ and $\mu_B$ shown in Table 1 have
been fitted with the $\chi^2$ method to the experimental data listed
in the third column. We stress a very good quality of the fit,
reflected by the low value of $\chi^2$ per degree of freedom. An
interesting feature of our fit is the stable value of the temperature,
when compared to other fits done at lower energies. The optimum value
of $T$ is practically the same as that found in the analysis of the
RHIC data at $\sqrt{s_{NN}}$= 130 GeV, and (within errors) it is the
same as the temperature found at the top-energy collisions at CERN SPS
\cite{mm}.\footnote{The similarities of the SPS and RHIC spectra 
are discussed in Ref.~\cite{spsrhic}.} On the other hand we observe a clear drop of the baryon
chemical potential, from 41~MeV found at $\sqrt{s_{NN}}$= 130~GeV down
to 29~MeV found in the present calculation. A smaller value of the
baryon chemical potential is a simple consequence of the observed
increase of the ${\bar p}/p$ ratio with the beam energy. At
mid-rapidity, the ratio asymptotes to unity with the increasing beam
energy.
We note the values of parameters shown in Table 1 are compatible within the statistical errors 
wit the results of Braun-Munzinger, Redlich, and Stachel \cite{review}, 
who obtain $T_{\rm chem}=177\pm7$~MeV and
$\mu_B=29\pm 6$~MeV. 

\begin{figure}[t]
\epsfysize=16.0cm
\par
\begin{center}
\mbox{\epsfbox{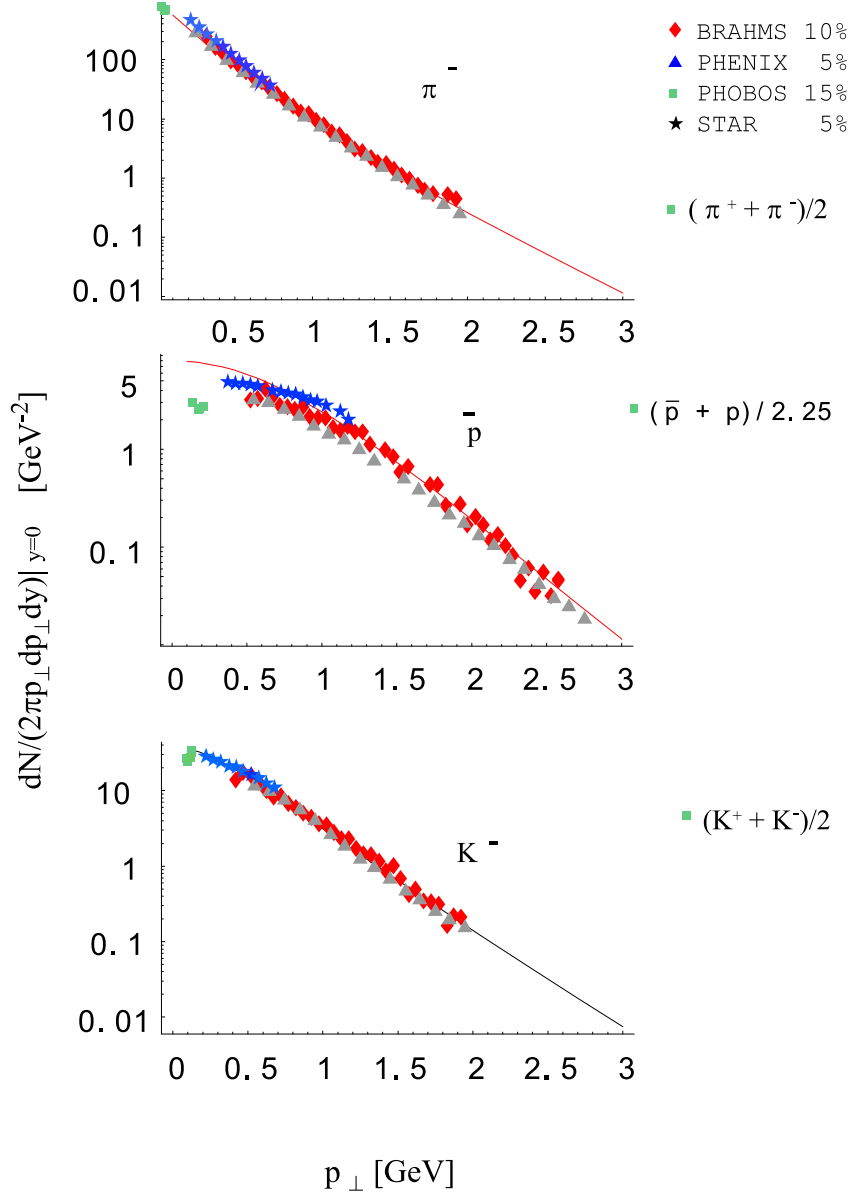}}
\end{center}
\caption{The transverse-momentum spectra of pions, protons, 
  and kaons measured by BRAHMS (diamonds), PHENIX (triangles), PHOBOS
  (squares), and STAR (stars) for the most central events in the
  $Au+Au$ collisions at $\sqrt{s_{NN}}=200$~GeV. The preliminary data are used in
  the form as compiled in Ref. \cite{ullrich}. The curves represent the
  result of the fit obtained in the single freeze-out model. Full feeding from 
weak decays is included in the calculation.}
\label{fig:BPPSmc}
\end{figure}

\section{Spectra}

Let us turn now to the study of the transverse-momentum spectra of
hadrons. They are calculated with the help of the Cooper-Frye
\cite{CF1,CF2} formula, 
\begin{equation}
\frac{dN}{d^{2}p_{\perp }dy} =
\int p^{\mu }d\Sigma _{\mu }\ f\left(p\cdot u\right) ,
\label{Ni}
\end{equation}
where the distribution function $f$ includes the products of the resonance
decays (more details on the construction of $f$ in the presence of the
resonance decays have been given in Refs.~\cite{str,zakop}).  The
element of the hypersurface, $d\Sigma_{\mu }$, is defined as
\begin{equation}
d\Sigma_\mu
= \epsilon_{\mu \alpha \beta \gamma}
{\partial x^\alpha \over \partial \alpha_\parallel}
{\partial x^\beta \over \partial \alpha_\perp}
{\partial x^\gamma \over \partial \phi} \, d\alpha_\parallel 
d\alpha_\perp d\phi, 
\label{sigma}
\end{equation}
where $x^0=t$, $x^1=r_x$, $x^2=r_y$, $x^3=r_z$, and $\epsilon_{\mu
  \alpha \beta \gamma}$ is the Levi-Civita tensor. Introducing a
convenient parameterization \cite{BL}:
\begin{eqnarray}
t &=&\tau \cosh \alpha _{\parallel }\cosh \alpha _{\perp },\quad r_{z}=\tau
\sinh \alpha _{\parallel }\cosh \alpha _{\perp },  \nonumber \\
r_{x} &=&\tau \sinh \alpha _{\perp }\cos \phi ,\quad r_{y}=\tau \sinh \alpha
_{\perp }\sin \phi ,  
\label{par}
\end{eqnarray}
one finds
\begin{equation}
d\Sigma^\mu(x) = u^\mu(x)\, \tau ^{3} \, {\rm sinh}(\alpha _{\perp})
{\rm cosh}(\alpha _{\perp}) \, d\alpha _{\perp}
d\alpha _{\parallel } d\phi,
\label{prop}
\end{equation}
hence the four-vectors $d\Sigma^\mu$ and $u^\mu$ are parallel. This
feature is special for our modeling of the freeze-out, 
see Eqs.~(\ref{tau}) and (\ref{umu}), and allows us to represent the spectra in
a compact form given by Eq. (\ref{Ni}). With the use of the
parameterization (\ref{par}) we further rewrite Eq. (\ref{Ni}) in the form
\begin{equation}
\frac{dN}{d^{2}p_{\perp }dy} =\ \tau ^{3}\int_{-\infty }^{+\infty
}d\alpha _{\parallel }\int_{0}^{\rho _{\max }/\tau }{\rm sinh}  \alpha _{\perp
}d\left( {\rm sinh}  \alpha _{\perp }\right) \int_{0}^{2\pi }
d\xi \, p\cdot u \, f\left( p\cdot u\right),
\label{dNi}
\end{equation}
where
\begin{equation}
p\cdot u=m_{\perp }{\rm cosh} \alpha _{\parallel } {\rm cosh}  \alpha
_{\perp }-p_{\perp }\cos \xi \, {\rm sinh}  \alpha _{\perp }. 
\label{pu}
\end{equation}
We note that $\alpha _{\parallel }$ is the rapidity of the fluid
element, $v_{z}=r_z/t=\tanh \alpha _{\parallel }$, and $\alpha
_{\perp}$ describes the transverse size of the system, $\rho =\tau
\sinh \alpha _{\perp}$.

The maximum and average
transverse-flow parameter is given in our model by the
equations
\begin{eqnarray}
\beta_\perp^{\rm max}= \frac{\rho_{\rm max}}{\sqrt{\tau^2+\rho_{\rm max}^2}}, \;\;\;
\langle \beta_\perp \rangle = \frac{\int_0^{\rho_{\rm max}} \rho d\rho 
\frac{\rho}{\sqrt{\tau^2+\rho^2}}}{\int_0^{\rho_{\rm max}} \rho d\rho} . \label{betaav}
\end{eqnarray}

\begin{figure}[t]
\epsfysize=14.3cm
\par
\begin{center}
\mbox{\epsfbox{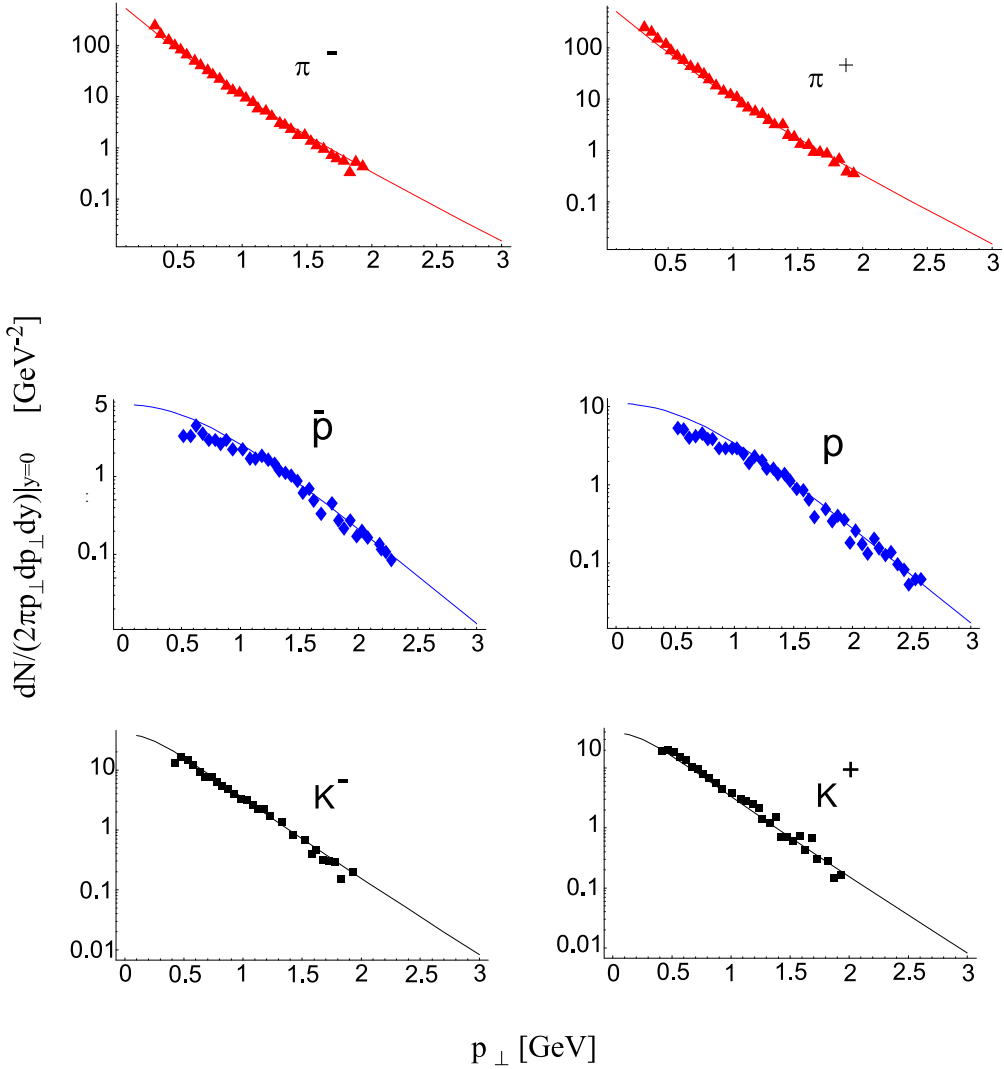}}
\end{center}
\caption{Our fit to the transverse-momentum 
  spectra of both the negative and positive hadrons measured by BRAHMS \cite{brahmsy}
  for most central events for the 
$Au+Au$ collisions at $\sqrt{s_{NN}}=200$~GeV. Full feeding from 
weak decays is included in the model calculation.}
\label{fig:brahm200}
\end{figure}

Since the ratios of hadron abundances depend  very weakly on the centrality
of the collision, we regard our thermodynamic parameters $T$ and
$\mu_B$ as universal. On the other hand, the two geometric parameters
may be different for each centrality class. From the obvious reasons
we expect that $\tau$ and $\rho_{\rm max}$ are larger for more central
events or, in other words, they are decreasing functions of the
centrality, $c$, defined by the formula
\begin{equation}
c={\pi b^2 \over \sigma_{\rm tot}},
\label{centr}
\end{equation} 
where $b$ is the impact parameter and $\sigma_{\rm tot}$ is the total inelastic cross section for
the colliding gold nuclei (for a discussion of the validity of this
simple geometric relation see Ref. \cite{centr}).  For various
centrality bins we fit the geometric parameters with the help of the
$\chi^2$ method, {\em i.e.} we minimize the expression
\begin{equation}
\chi^{2}(\tau,\rho_{\rm max})=\sum_{n=1}^{n_{\rm max}}
{\left[D_{n}^{\rm exp}-D_{n}^{\rm model}(\tau,\rho_{\rm max})\right]^{2} 
\over \sigma_{n}^{2}},
\label{chi2}
\end{equation} 
where $D_{n}^{\rm exp}$ is the n-th measured value of the
transverse-momentum spectrum, $D_{n}^{\rm model}(\tau,\rho_{\rm max})$
is the corresponding value calculated in our model, and $\sigma_{n}$
is the error.  Observing the differences between the points measured
by different experimental groups we conclude that the systematic
errors are much larger than the statistical errors. Therefore, we
assume that $\sigma_{n}$ is determined by the systematic
uncertainties and take $\sigma_{n}=D_{n}^{\rm exp}/10$.\footnote{
We note that the $\chi^2$ method was also used to determine the values
of the two thermodynamic parameters in Sect. 3. Since the thermodynamic 
parameters are sensitive only to the ratios of hadron abundances our
fitting of the four parameters $T$, $\mu_B$, $\tau$, and $\rho_{\rm max}$
can be done in two steps: first we determine  $T$ and $\mu_B$, and
having fixed these two values we determine later $\tau$ and $\rho_{\rm max}$.
One could follow also a different path by fitting simultaneously the four
parameters to the experimental spectra.}

\begin{figure}[t]
\epsfysize=10.0cm
\par
\begin{center}
\mbox{\epsfbox{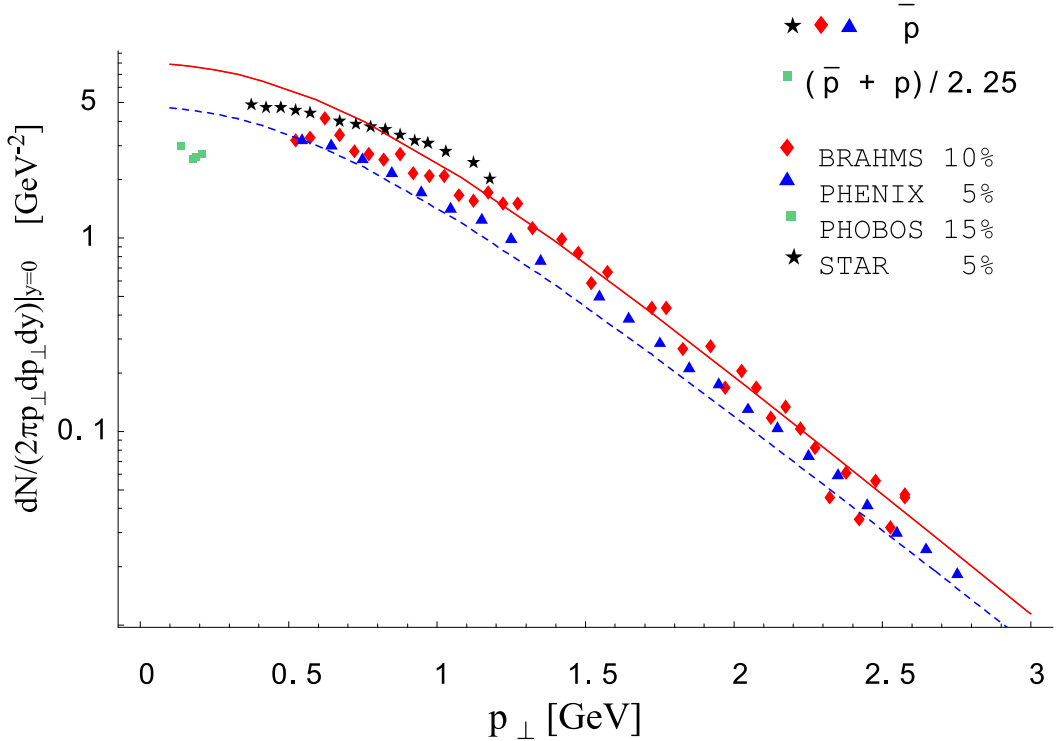}}
\end{center}
\caption{Our fit to the transverse-momentum spectra of antiprotons
  measured by the four RHIC experimental groups for the most central events
of the $Au+Au$ collisions at $\sqrt{s_{NN}}=200$~GeV. The preliminary data are used in
  the form as compiled in Ref. \cite{ullrich}. 
  The solid curve describes the model result with full feeding of
  antiprotons from the weak decays, whereas the dashed line describes
  the model calculation without any contributions from the weak decays.}
\label{fig:pbar200}
\end{figure}

\section{Most central events}

In Fig \ref{fig:BPPSmc} we show the result of our fit to the
transverse-momentum spectra of $\pi^-,\,{\bar p}$ and $K^-$, for the
most central events measured by the four RHIC experiments
\footnote{Most of the data shown in our figures 
were digitized from the available plots with preliminary 
experimental results.}
(exceptionally, the data of PHOBOS are for the mixtures of negative
and positive hadrons). In this case, the centrality class of the
BRAHMS and PHOBOS data is 10\%, whereas the centrality class of the
PHENIX and STAR data is 5\%. Our model curves for $\pi^-,\,{\bar p}$
and $K^-$, shown in Fig.  \ref{fig:BPPSmc}, are simultaneous fits to
the four experimental spectra available for each hadron species. The
fitted values of the geometric parameters are in this case $\tau$ =
7.58 fm and $\rho_{\rm max}$ = 7.27 fm (see Table 2 for a complete
list of the values of the geometric parameters found for different
experimental situations).  We observe that these numbers are similar
to those found in our previous study of the RHIC data collected at a
lower energy of $\sqrt{s_{NN}}$ = 130~GeV, $\tau$ = 7.66~fm and
$\rho_{\rm max}$ = 6.69~fm \cite{wbwf,str}. This weak dependence of
our geometric parameters on the energy of the colliding ions is
reminiscent of the very weak energy dependence of the measured
transverse HBT radii $R_{\rm side}$ and $R_{\rm out}$
\cite{phenixhbt,starhbt,agshbt,coimbrab}.

In Fig. \ref{fig:brahm200} we show our fit to the most central BRAHMS
data only \cite{brahmsy}. In this case the fitted values of the geometric parameters
are $\tau = 7.68$~fm and $\rho_{\rm max} = 7.46$~fm. These values are
consistent with the previous values obtained from the ``mixed'' input.
In Figs. \ref{fig:BPPSmc} and \ref{fig:brahm200} we can see that our
model curves describe the data very well in the large range of
$p_\perp$ stretching up to about 2~GeV.

\begin{figure}[tb]
\epsfysize=16.5cm
\par
\begin{center}
\mbox{\epsfbox{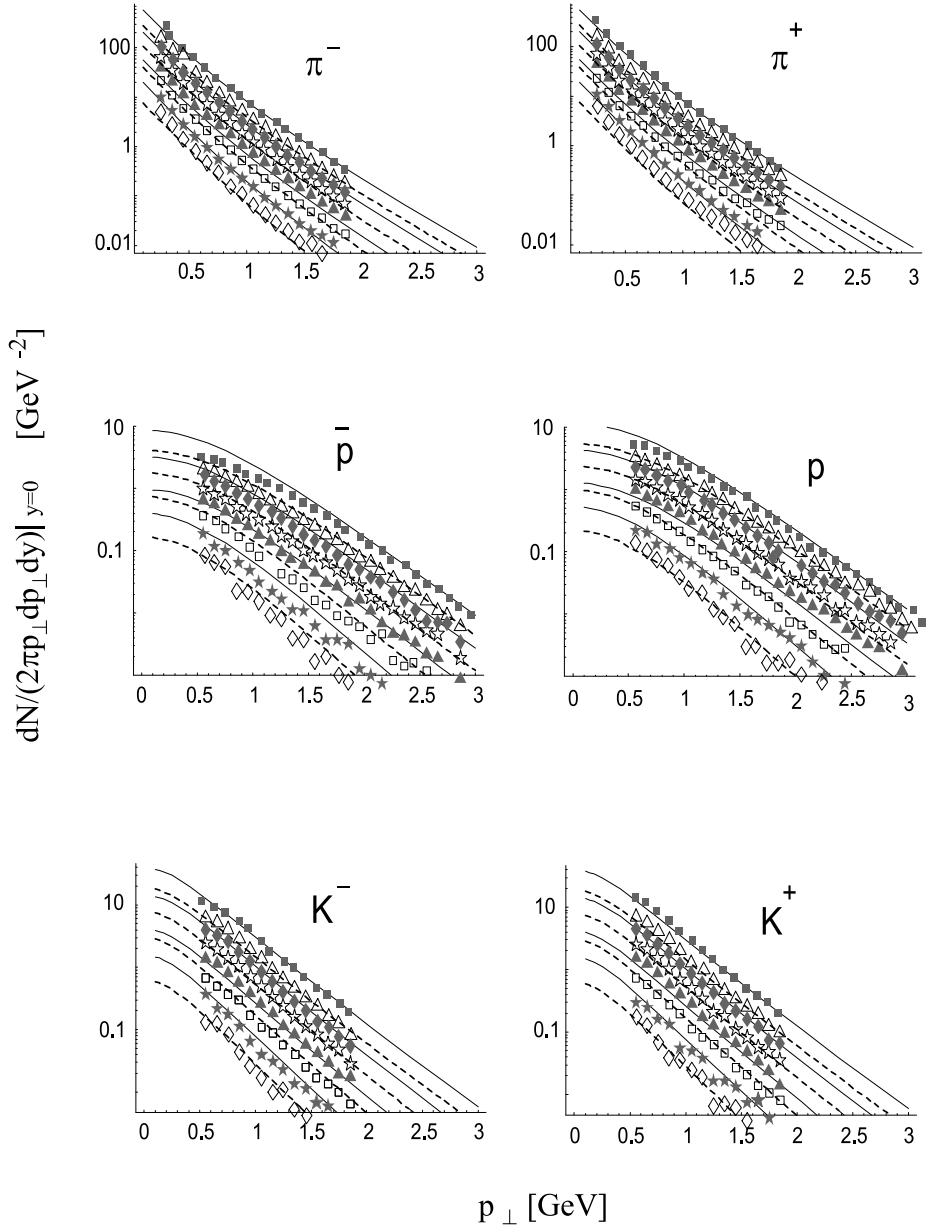}}
\end{center}
\caption{Our fit to the transverse-momentum spectra of both the negative 
and positive hadrons measured by 
  PHENIX \cite{chujo} for the $Au+Au$ collisions at $\sqrt{s_{NN}}=200$~GeV.
The eight sets of points and the fitted curves correspond to the eight centrality bins 
given in Table 2. The two thermal parameters have uniform values for all cases, while the 
geometric parameters are adjusted independently for each centrality bin, with results 
displayed in Table 2. Full feeding from 
weak decays is included in the model calculation.}
\label{fig:ph200}
\end{figure}

A closer inspection of Figs.~\ref{fig:BPPSmc} and \ref{fig:brahm200}
shows that our model curve tends to overestimate the spectra of
antiprotons at small values of $p_\perp$. To study the possible origin
of this effect, in Fig. \ref{fig:pbar200} we show in detail the
antiproton spectra measured by different experimental groups. One can
see that the experimental spectra differ substantially from each other
and, in practice, one cannot fit all four groups of the experimental
points simultaneously. The most likely reason for such differences are
difficulties connected with the correct estimate of the role of the
weak decays, or with fixing the normalization. Our model calculations
shown in Fig.~\ref{fig:pbar200} indicate that the effect of the weak
decays on the antiproton spectrum is indeed important. The solid line
represents our result obtained in the case where the antiprotons are
fed by the weak decays.  This curve may be treated as the upper
limit for the spectrum.  On the other hand, the dashed line represents
our result with no feeding included,
and may be treated as the lower
limit. We can see that most of the experimental points, except for those
of PHOBOS, lie in the band enclosed with our two curves.

\begin{figure}[t]
\epsfysize=9.7cm
\par
\begin{center}
\mbox{\epsfbox{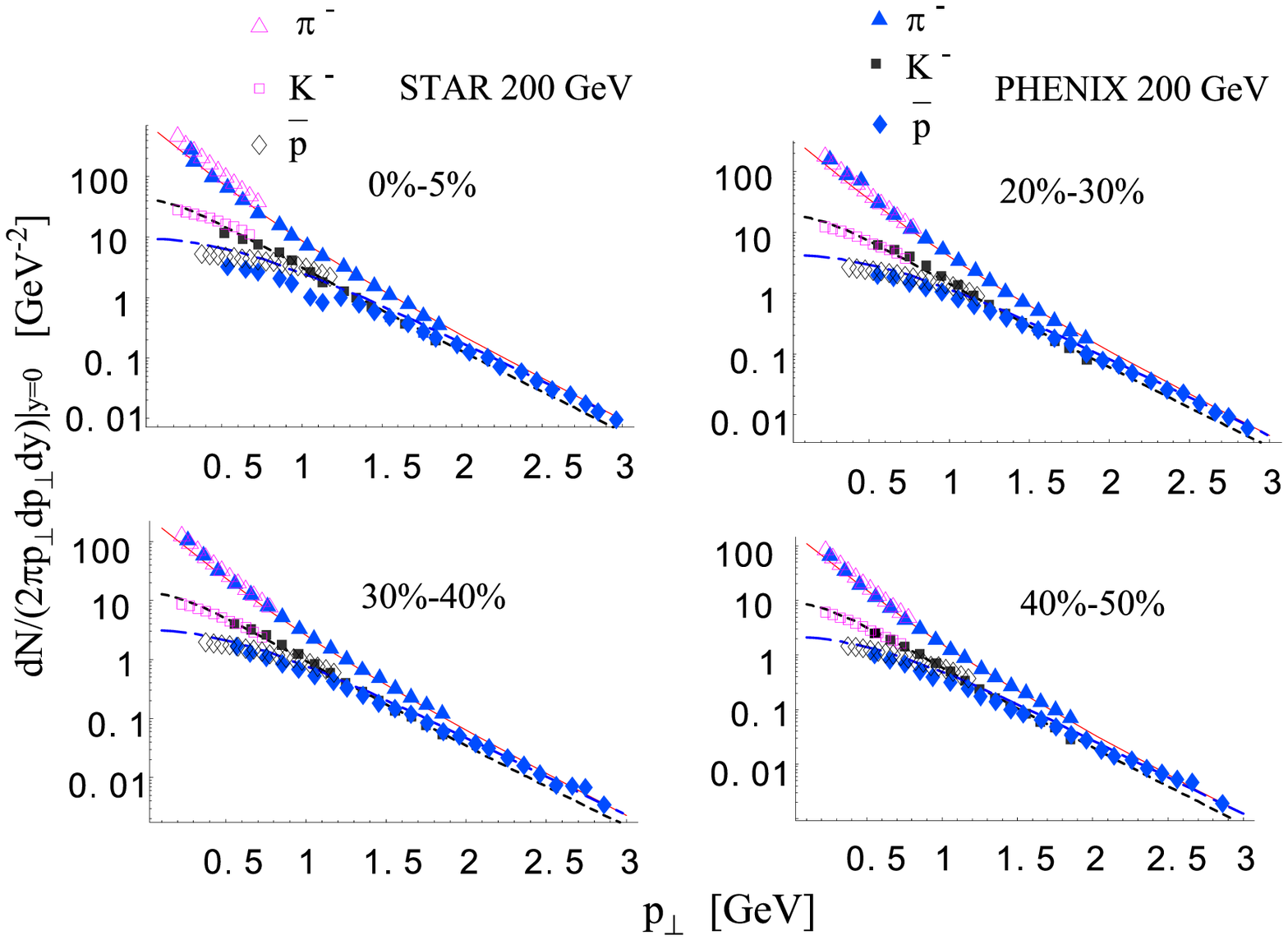}}
\end{center}
\caption{Our joint fit to the transverse-momentum spectra from PHENIX \cite{chujo} 
and STAR \cite{star200}: 
  pions (triangles), kaons (squares), and antiprotons measured by STAR
  (open symbols) and PHENIX (filled symbols) for the $Au+Au$ collisions
  at $\sqrt{s_{NN}}=200$~GeV. Full feeding from 
weak decays is included in the model calculation.}
\label{stiphen200}
\end{figure}

\begin{table}
\begin{center}
\begin{tabular}{|l|c|c|c|c|c|}
\hline
        & $c$ [\%] & $\tau$ [fm]       & $\rho_{\rm max}$ [fm] & $\langle \beta_\perp \rangle$ & $\beta_\perp^{\rm max}$\\
\hline
PHENIX +  & $0-5$  & $7.58 \pm 0.32$   & $7.27 \pm 0.12$ & $0.52 \pm 0.02 $ & $0.69 \pm 0.02$  \\
STAR +    & $0-5$  &                   &                 & &  \\
PHOBOS +  & $15$   &                   &                 & &  \\
BRAHMS    & $10$   &                   &                 & &  \\
\hline
BRAHMS  
& $10$     & $7.68 \pm 0.19$   & $7.46 \pm 0.05$ & $0.52 \pm 0.01 $ & $0.70 \pm 0.01$    \\                                          
\hline
STAR    
& $0-5$    & $9.74 \pm 1.57$   & $7.74 \pm 0.68$ & $0.45 \pm 0.08 $ & $0.62 \pm 0.09$ \\  
& $5-10$   & $8.69 \pm 1.39$   & $7.18 \pm 0.64$ & $0.47 \pm 0.08 $ & $0.64 \pm 0.09$ \\
& $10-20$  & $8.12 \pm 1.31$   & $6.44 \pm 0.57$ & $0.45 \pm 0.08 $ & $0.62 \pm 0.10$ \\
& $20-30$  & $7.24 \pm 1.18$   & $5.57 \pm 0.50$ & $0.44 \pm 0.08 $ & $0.61 \pm 0.10$ \\
& $30-40$  & $7.07 \pm 1.17$   & $4.63 \pm 0.39$ & $0.39 \pm 0.08 $ & $0.55 \pm 0.10$ \\
& $40-50$  & $6.38 \pm 1.02$   & $3.91 \pm 0.33$ & $0.37 \pm 0.07 $ & $0.52 \pm 0.09$ \\
& $50-60$  & $6.19 \pm 1.09$   & $3.25 \pm 0.28$ & $0.32 \pm 0.07 $ & $0.46 \pm 0.10$ \\
& $70-80$  & $5.48 \pm 0.81$   & $4.03 \pm 0.10$ & $0.43 \pm 0.06 $ & $0.59 \pm 0.07$ \\
\hline
PHENIX   
& $0-5$    & $7.86 \pm 0.38$   & $7.15 \pm 0.13$ & $0.50 \pm 0.02 $ & $0.67 \pm 0.02$  \\
& $20-30$  & $6.14 \pm 0.32$   & $5.62 \pm 0.11$ & $0.50 \pm 0.02 $ & $0.68 \pm 0.03$  \\
& $30-40$  & $5.73 \pm 0.16$   & $4.95 \pm 0.05$ & $0.48 \pm 0.01 $ & $0.65 \pm 0.01$  \\
& $40-50$  & $4.75 \pm 0.28$   & $3.96 \pm 0.09$ & $0.47 \pm 0.03 $ & $0.64 \pm 0.03$  \\
& $50-60$  & $3.91 \pm 0.23$   & $3.12 \pm 0.07$ & $0.45 \pm 0.03 $ & $0.62 \pm 0.03$  \\
& $60-70$  & $3.67 \pm 0.12$   & $2.67 \pm 0.03$ & $0.42 \pm 0.01 $ & $0.59 \pm 0.02$  \\
& $70-80$  & $3.09 \pm 0.11$   & $2.02 \pm 0.02$ & $0.39 \pm 0.01 $ & $0.55 \pm 0.02$  \\
& $80-91$  & $2.76 \pm 0.20$   & $1.43 \pm 0.03$ & $0.32 \pm 0.03 $ & $0.46 \pm 0.03$  \\                         
\hline
PHENIX + STAR   
& $0-5$    & $7.96 \pm 0.41$  & $7.15 \pm 0.17$  & $0.50 \pm 0.03 $ & $0.67 \pm 0.03$ \\
& $20-30$  & $6.08 \pm 0.32$  & $5.53 \pm 0.13$  & $0.50 \pm 0.03 $ & $0.67 \pm 0.03$ \\
& $30-40$  & $5.59 \pm 0.30$  & $4.75 \pm 0.11$  & $0.48 \pm 0.03 $ & $0.65 \pm 0.03$ \\
& $40-50$  & $4.95 \pm 0.27$  & $4.03 \pm 0.10$  & $0.46 \pm 0.03 $ & $0.63 \pm 0.03$ \\
\hline                          
\end{tabular}
\caption{The values of the optimum geometric parameters, $\tau$ and $\rho_{\rm max}$, and the 
corresponding average and maximum transverse flow velocities, 
$\langle \beta_\perp \rangle$ and $\beta_\perp^{\rm max}$ of Eq.~(\ref{betaav}), obtained 
  from our analysis of the data collected by different experimental
  groups at different values of the centrality. All results are for the
  beam energy of $\sqrt{s_{NN}}=$ 200 GeV. The model calculation for STAR includes the corrections
for the weak decays in a way used by the STAR Collaboration, {\em i.e.}, the feeding of the pions 
from the decays of $\Lambda$ is excluded. The model calculations for BRAHMS and PHENIX 
include full feeding from the weak decays.}
\label{invsl}
\end{center}
\end{table}

\section{Centrality dependence of the geometric parameters}

Having at our disposal the data collected in different centrality
windows we may determine the dependence of the expansion parameters on
$c$, or, equivalently, on the impact parameter $b$. In Fig. \ref{fig:ph200} we show
the PHENIX data together with our model curves. Both the data and the
model spectra are plotted for eight different centrality classes. The most
central collision correspond to the centrality class $c=$ 0-5\%,
whereas the most peripheral collisions included in Fig.
\ref{fig:ph200} are for the centrality class $c=$ 80-91\%. The values
of the fitted expansion parameters for all eight cases are given in Table~2.

The geometric parameters obtained from the fit to the STAR data turn
out to be noticeably larger compared to the parameters obtained from
the analysis of the PHENIX data. In order to see the origin of this effect, in
Fig. \ref{stiphen200} we superimpose the spectra from STAR and PHENIX
for four centrality classes common for the two experiments. We can see
that the spectra measured by STAR (open symbols) are lying slightly
above the spectra measured by PHENIX (filled symbols). As a consequence
of this behavior, our normalization-controlling parameter, $\tau$, turns
out to be larger for STAR. We have performed a simultaneous fit to
the PHENIX and STAR data. The model curves for this case are also
shown in Fig.  \ref{stiphen200} (solid, dashed, and long-dashed
curves, respectively for pions, kaons, and antiprotons). The fitted
values of the expansion parameters $\tau$ and $\rho_{\rm max}$ are
listed in the lowest entry of Table 2.

\begin{figure}[h]
\epsfysize=6.0cm
\par
\begin{center}
\mbox{\epsfbox{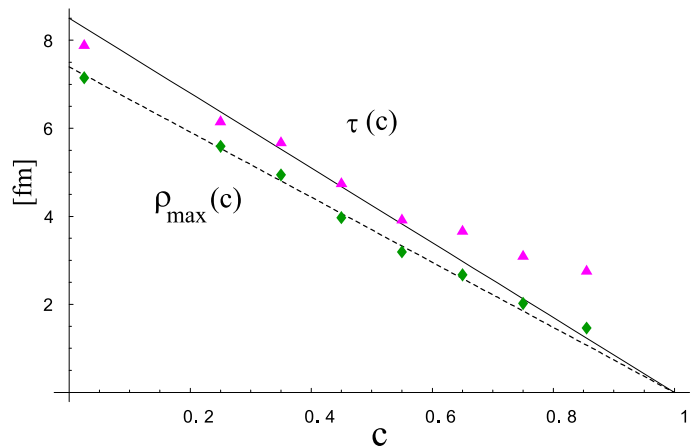}}
\end{center}
\caption{Centrality dependence of the geometric parameters, $\tau$ and $\rho_{\rm max}$,
  as extracted from the analysis of the PHENIX data only.}
\label{fig1}
\end{figure}

\begin{figure}[h]
\epsfysize=6.0cm
\par
\begin{center}
\mbox{\epsfbox{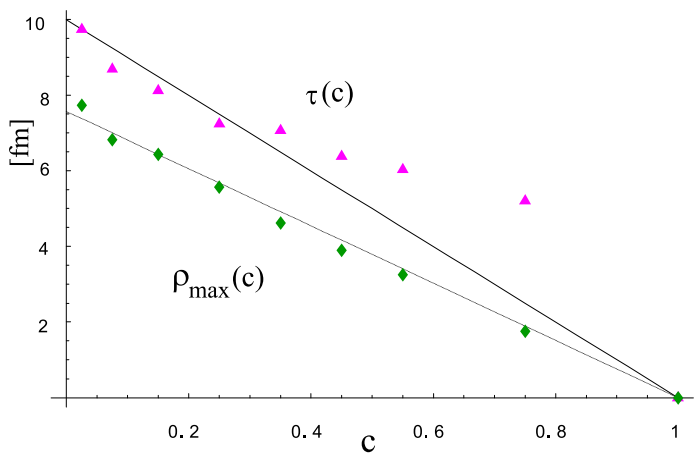}}
\end{center}
\caption{Centrality dependence of the geometric parameters $\tau$ and $\rho_{\rm max}$ 
  as extracted from the analysis of the STAR data only.}
\label{fig2}
\end{figure}

An interesting behavior can be found when the values of $\tau$ and
$\rho_{\rm max}$ are plotted versus $c$ (see Figs.
\ref{fig1} and \ref{fig2}). The parameter $\rho_{\rm max}$ exhibits
almost ideal linear dependence on $c$ in the full range $0<c<1$.
Similarly, the parameter $\tau$ decreases linearly with growing $c$ in
the range $0 < c < 0.5$.  For larger values of $c$ the drop of $\tau$
is slightly weaker, although the data of PHENIX show a rather strong
tendency to line up.  The approximate linear scaling of both $\tau$
and $\rho_{\rm max}$ with $c$ has an important physical significance.
The hadron multiplicities in our model are functions of $\tau^3$ and
$\rho_{\rm max}/\tau$, see Eq. (\ref{dNi}), hence the linear dependence of
$\tau$ and $\rho_{\rm max}$ on $c$ means that the ratio $\rho_{\rm
  max}/\tau$ is independent of $c$ and we are left with the dependence
on $\tau^3$ only. Consequently, using Eq. (\ref{centr}) we find that
the total multiplicity in the central region (mid-rapidity)
obtained in our model should, to a good approximation,
scales in the following way
\begin{equation}
N_{\rm model}(b) = N_{\rm model}(b=0) \left(1-c(b)\right)^3 =  
N_{\rm model}(b=0) \left(1-{\pi b^2 \over \sigma_{\rm tot} }\right)^3.
\label{eq:scale}
\end{equation}
This type of the behavior may be compared to the $b$-dependence of
hadron production following from the wounded nucleon model
\cite{wounded}. For the symmetric case, {\em i.e.} when two identical
nuclei collide, the number of the wounded nucleons is given by the
formula
\begin{figure}[tb]
\epsfysize=9.0cm
\par
\begin{center}
\mbox{\epsfbox{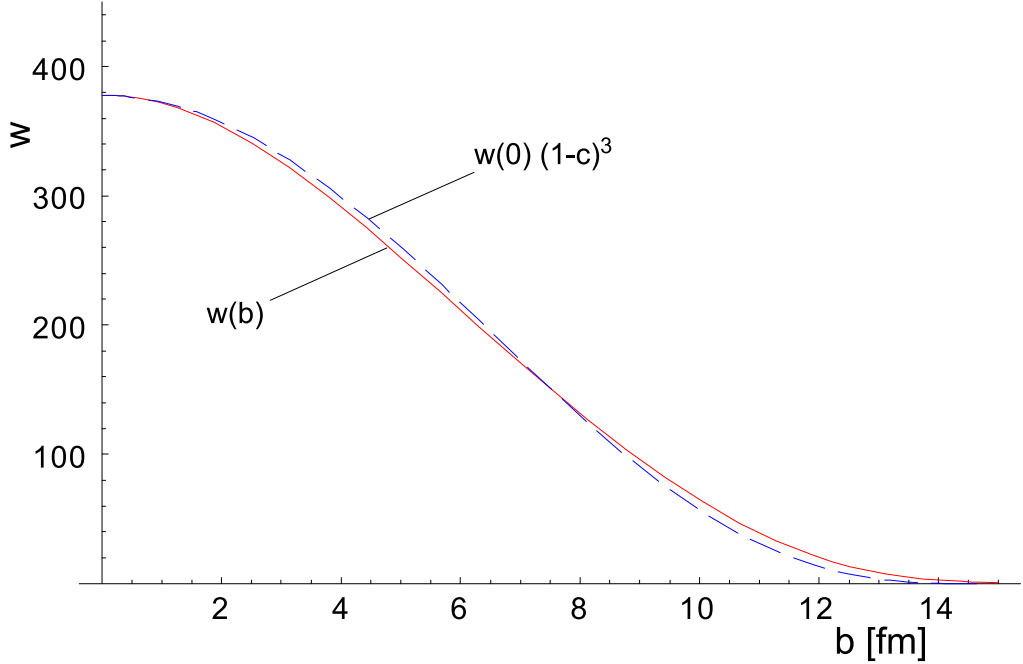}}
\end{center}
\caption{The number of wounded nucleons, $w(b)$ (solid line)
and the approximating function $w(0)(1-c(b))^3$ (dashed line), plotted as functions of the 
impact parameter $b$. Since the multiplicity of hadrons produced in our model is 
proportional to $(1-c(b))^3$ at moderate values of $c$, the model conforms to the 
wounded-nucleon scaling.}
\label{scale}
\end{figure}
\begin{equation}
w\left(b\right)=
2 A\int d^{2}s\,T_{A}\left( {\bf b}-{\bf s}\right) \left( 1-\left[ 1-\sigma
_{\rm {in}}T_{A}\left( {\bf s}\right) \right] ^{A}\right)
\label{wound}
\end{equation}
where $A$ is the mass number, $\sigma_{\rm in}$ is the inelastic
nucleon-nucleon cross section, and $T_A$ is the nucleon-nucleus
thickness function. Similarly, the total inelastic cross section is given by the 
formula
\begin{equation}
\sigma _{\rm {tot}}=\int d^{2}b\left(1-\left[1-\sigma_{\rm {in}}  T_{AA}\left({\bf b}\right)
\right] ^{A^2}\right),  
\label{sigtot}
\end{equation}
where $T_{AA}\left({\bf b}\right)$ is the nucleus-nucleus thickness function.
Using the values $A=197$ and $\sigma_{\rm in}$=40 mb, and assuming the
Woods-Saxon distribution for the nuclear densities (with the standard
choice of the parameters) we find
\begin{equation}
w\left(b\right) \approx w(0)(1-c(b))^3.
\label{wofb}
\end{equation}
Since the wounded-nucleon model assumes the scaling of hadron
production with the number of the wounded nucleons, the total
multiplicity of hadrons considered in this model scales also as
$(1-c(b))^3$,
hence it is compatible with the scaling obtained in our
approach, {\em cf.} Figs.~\ref{fig1} and \ref{fig2}.  
The quality of the simple approximation (\ref{wofb}) is
shown in Fig.  \ref{scale}.  This result means that our model, with
the geometric parameters fit to reproduce the data at various
centralities, conforms with high accuracy to the wounded-nucleon
scaling for not-too-large values of $c$ (say, $c<0.5$).

We end this Section with a technical remark.  In the analysis of this
Section we have ignored any possible effects of the azimuthal
deformation.  Clearly, the emitting source formed in a non-central
collision will not be azimuthally-symmetric, which, by the way, is
seen in the HBT measurements \cite{HBTazim}. However, it turns out
that the effects of this deformation are tiny for the $p_T$-spectra
averaged over the azimuthal angle \cite{ABaran}, thus can be dropped
in the analysis of the present paper.

\begin{figure}[b]
\epsfysize=7.7cm
\par
\begin{center}
\mbox{\epsfbox{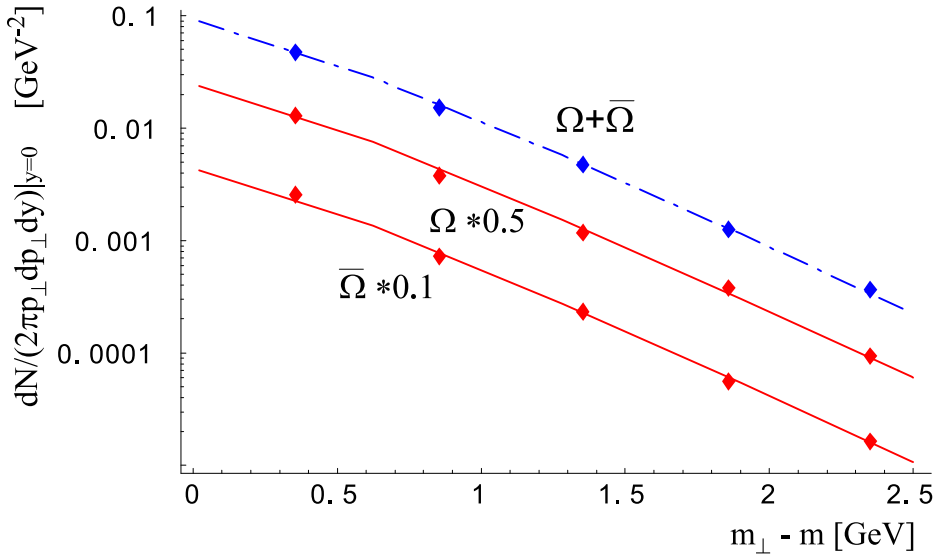}}
\end{center}
\caption{Our fit to the transverse-momentum spectra of $\Omega$ and $\bar \Omega$, 
compared 
  measured by the four RHIC experimental groups for the most central events
of the $Au+Au$ collisions at $\sqrt{s_{NN}}=200$~GeV. The preliminary data are taken 
from Ref. \cite{suire}. The values of the geometric parameters used in the model calculation are 
$\tau=8.3$~fm and $\rho_{\rm max}=7.1$~fm.}
\label{fig:omega}
\end{figure}

\section{Spectrum of the $\Omega$ baryons}

Finally, we wish to present the $p_\perp$ spectra for the $\Omega$ and $\bar \Omega$.
Similarly to the case  of the lower energy of $\sqrt{s_{NN}}=130$~GeV \cite{str}, our model is in full
agreement with the preliminary data of Ref.~\cite{suire}, which can be seen in Fig.~\ref{fig:omega}. 
In the model calculation we have used the thermal parameters from Table 1 and the geometric 
parameters $\tau=8.3$~fm and $\rho_{\rm max}=7.1$~fm, which roughly corresponds to 
centrality $0-10$\% in the STAR experiment ({\em cf.} Table 2).

\section{Conclusions}

It is clear from the results presented in this paper that the single
freeze-out model works well for the particle ratios and the
transverse-momentum spectra for the full RHIC energy of
$\sqrt{s_{NN}}=200$~GeV. This confirms our results published
previously for $\sqrt{s_{NN}}=130$~GeV \cite{wbwf,str,zakop}.  The model provides
an economic framework to parameterize the data, and supports the use of
the thermal approach to heavy-ion physics. In addition, we have shown
that the model works equally well for the central, as well as for the
non-central collisions. The transverse-size parameter, 
$\rho_{\rm max}$, drops linearly practically for all values of
$c$, while the ratio of $\rho_{\rm max}$ to $\tau$ remains almost
constant for not-too-large values of $c$. The systematics of the
dependence of the geometric/flow parameters on the centrality is most
interesting. For moderate values of the parameter $c$, say $c<0.5$,
the analysis of the non-central collisions displays scaling in
accordance to the wounded-nucleon model.

\section*{Acknowledgements}
We are grateful to Fuqiang Wang and Patricia Fachini for many helpful 
remarks concerning the data.

\end{document}